# Channel Multiplexing in Wireless Terahertz Communications Using Orbital Angular Momentum States

**KATHIRVEL NALLAPPAN[1], HICHEM GUERBOUKHA[1], MOHAMED SEGHILANI[2], JOSÉ AZANA[2], CHAHÉ NERGUIZIAN[1], AND MAKSIM SKOROBOGATIY[1,*]**

[1]*École Polytechnique de Montréal, Montréal, Québec, H3T 1J4, Canada*
[2]*Institute National de la Recherche Scientifique, Montréal, Québec, H5A 1K6, Canada*
**Corresponding author: maksim.skorobogatiy@polymtl.ca*

**We present design and experimental validation of the system for the generation of the Orbital Angular Momentum (OAM) states using 3D-printed low-loss metamaterial phase plates for application in the terahertz (THz) wireless communications. By azimuthally varying the hole pattern density within the phase plate, the local effective refractive index is varied, thus also changing the local propagation constant in the azimuthal direction. The OAM of any topological charge can be created by simply varying the thickness of the phase plate. The phase plate with topological charge (m=1) is 3D printed and the amplitude and the phase of the terahertz signal after passing the plate is characterized using the THz-time domain imaging system. Finally, we present the experimental setup and theoretical simulation on the multiplexing and de-multiplexing of several different OAM states for applications in wireless terahertz communication.**

Space division multiplexing (SDM) is continuously growing in order to increase the information capacity and the spectral efficiency of the communication channel. One among the SDM technique is the multiplexing using orbital angular momentum (OAM) [1, 2] which increases the transmission capacity by a factor equal to the number of transmitted spatial modes [3]. An electromagnetic wave carrying OAM has a helical transverse phase in the form of exp(im$\phi$), where 'm' is the charge of the OAM and '$\phi$' is the azimuthal angle. The value of 'm' is either positive or negative which represents the number of $2\pi$ changes in the azimuthal direction [4]. OAM multiplexing can be added to the existing multiplexing techniques such as wavelength division multiplexing and polarization division multiplexing [5]. Other than using OAM in free-space optical communication (FSO), it can be employed in the line of sight (LOS) microwave and millimeter wave link for the transmission of multiple data channels through a single aperture [6-8]. The effects of other channel conditions beyond LOS such as multipath and object obstructions have also been studied [9]. It was shown that the multipath propagation due to reflections distorts the OAM beams resulting in the intra and inter-channel crosstalk, thereby degrading the link performance [10]. Several techniques have been used to generate the OAM beams in the radio frequencies. The OAM beams at 90 GHz were generated using spiral phase plate [11, 12]. A conventional phase array antennas were used to generate OAM by configuring each antenna element with an appropriate phase delay [13]. This technique can also be used simultaneously in the generation of OAM and beam steering [14]. Radially symmetric airy beams are also demonstrated recently to carry the OAM charge in the terahertz region [15].

In this article, we present the design for the generation of OAM using patterned phase plate in the THz region (100 GHz-10 THz). We also show the theoretical simulation on de-multiplexing the different OAM states spatially in real-time THz wireless communication experiements. The pattern phase plate is printed using the 3D stereo-lithography technique (Asiga) with PlasCLEAR as the photoresist material.

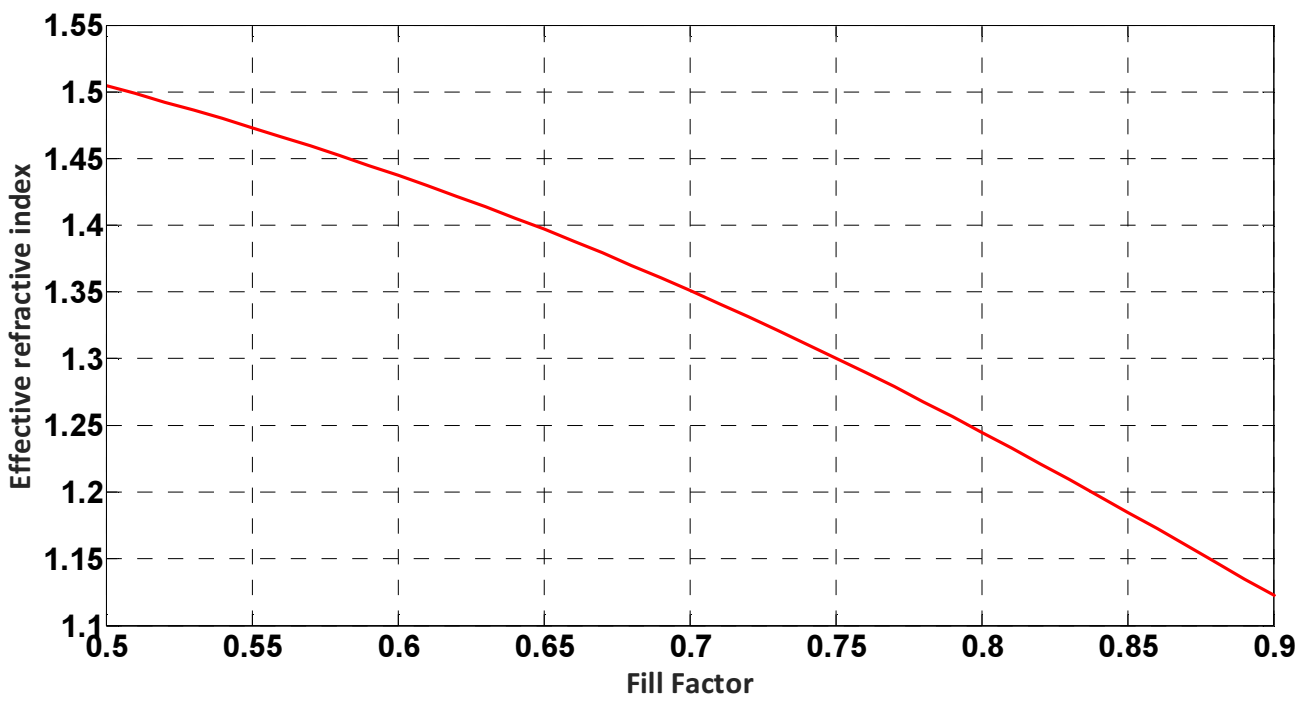

Fig.1. The variation of effective refractive index as a function of fill factor for the patterned metamaterial

The refractive index and the loss of the material is ~1.6 and 0.5 $cm^{-1}$ at 100 GHz respectively [16]. By azimuthally varying the hole size of the phase plate, the effective refractive index is varied and thereby changing the local propagation constant in the azimuthal direction. The effective refractive index is proportional to the fill factor (hole diameter/grating period). The patterned phase plate is designed with the fill factor varied from 0.5 to 0.9 which results in the effective index variation from 1.504 to 1.123 as shown in fig.1. The printed phase plate is soaked in the beaker containing isopropanol for a period of 5 hours to remove the residual photoresist and then polished to get the smooth surface. The polished phase plate is then UV cured for 2 minutes using Pico Flash (Asiga). The printed design and the UV cured phase plate are shown in fig.2. (a) and (b) respectively.

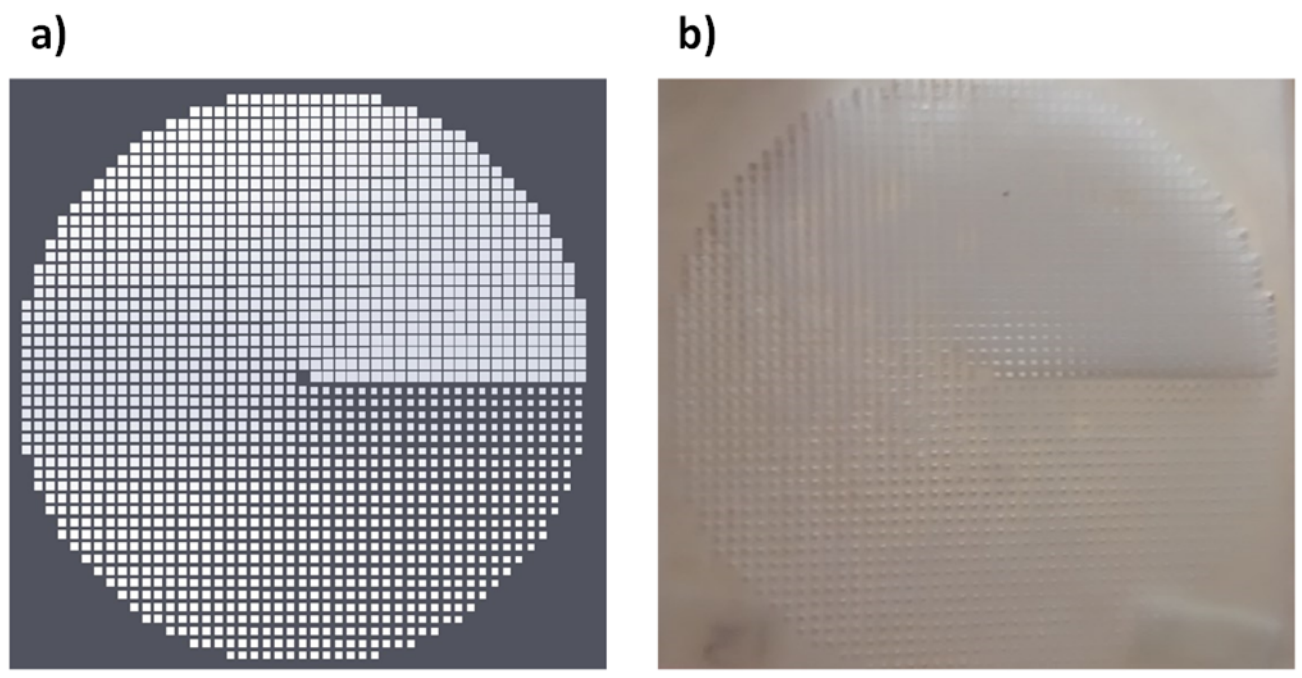

Fig.2. (a) The proposed design of patterned phase plate and (b) The 3D printed phase plate after UV curing.

The phase plate is characterized using the THz-Time domain system. The schematic of the imaging set up is shown in fig.3. A high power interdigitated antenna is used as the THz emitter and a photoconductive antenna is used as the THz detector. The emitter antenna is excited by the free space coupling of femtosecond laser whereas, the detector antenna is fiber coupled. In order to compensate the pulse broadening due to the dispersion in the fiber, a grating-based dispersion-compensation system is used before coupling the femtosecond pulse into the polarization maintaining fiber [17].

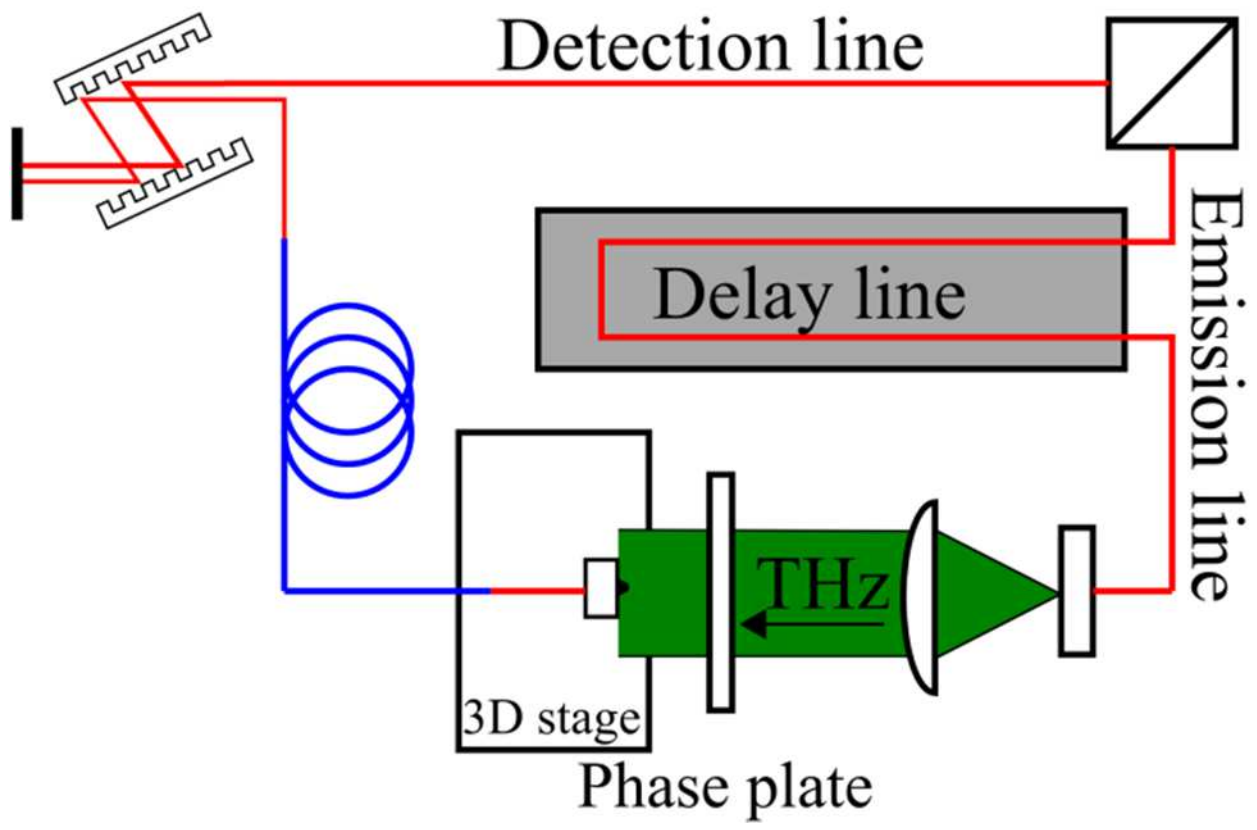

Fig.3. Schematic of the THz-Time domain system for characterizing the Phase plate. The detector antenna is mounted on a 3D stage and phase plate is raster scanned to record the THz beam profile after passing through the phase plate.

A Teflon lens is used to collimate the emitted THz wave and the phase plate is aligned perpendicular to the THz wave propagation. The detector antenna is mounted on a 3D stage and the phase plate is raster scanned with a step resolution of 1 mm to obtain the THz beam profile as shown in fig.4.

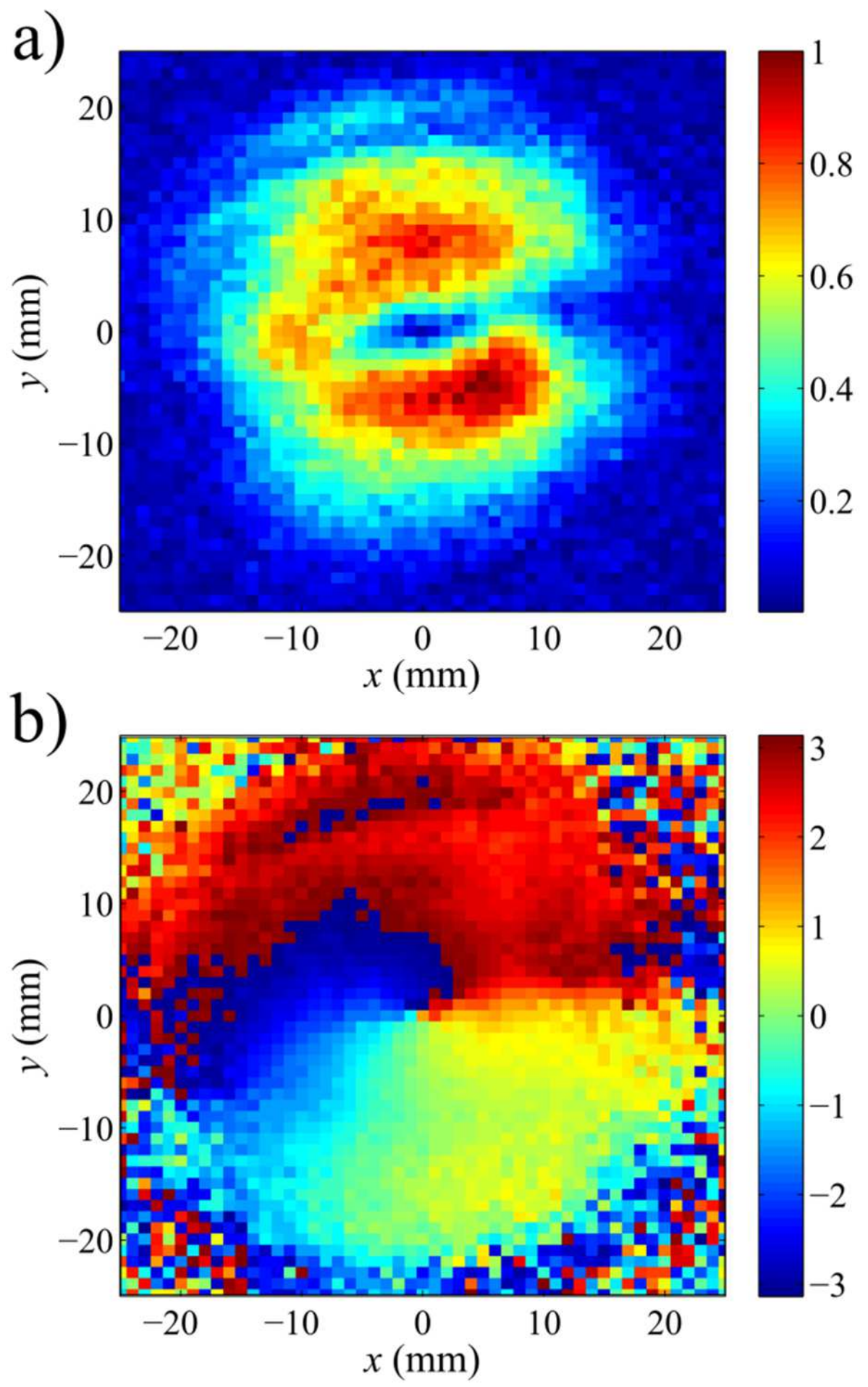

Fig.4. (a) Amplitude (b) phase image of the patterned phase plate at 100 GHz.

The donut shape of the amplitude image from fig.4. (a) confirms the conversion of a Gaussian beam into OAM beam. Fig.4. (b) shows the smooth 2π delay in the phase profile confirming the generation of OAM.

The OAM beam divergence is theoretically calculated in order to use the patterned phase plate in the real-time wireless communication for multiplexing and de-multiplexing respectively. For that the Gaussian beam is expressed in terms of Bessel beams as they are the Eigen states of the solution of Maxwell equations in free space. A linearly polarized Gaussian beam excites the Angular momentum of state m=+1 and m=-1 in the free space. After passing through the phase plate of topological charge m=1, two states of Angular momentum, m=0 and m=2 is excited. The excited states will have different divergence in the free space which will be used for de-multiplexing the channels in space domain. The simulation of beam divergence of the sum of linearly polarized field components at 140 GHz carrier frequency as a function of propagation distance (X-direction) is shown in fig.5, where $\omega_0$ is the beam waist.

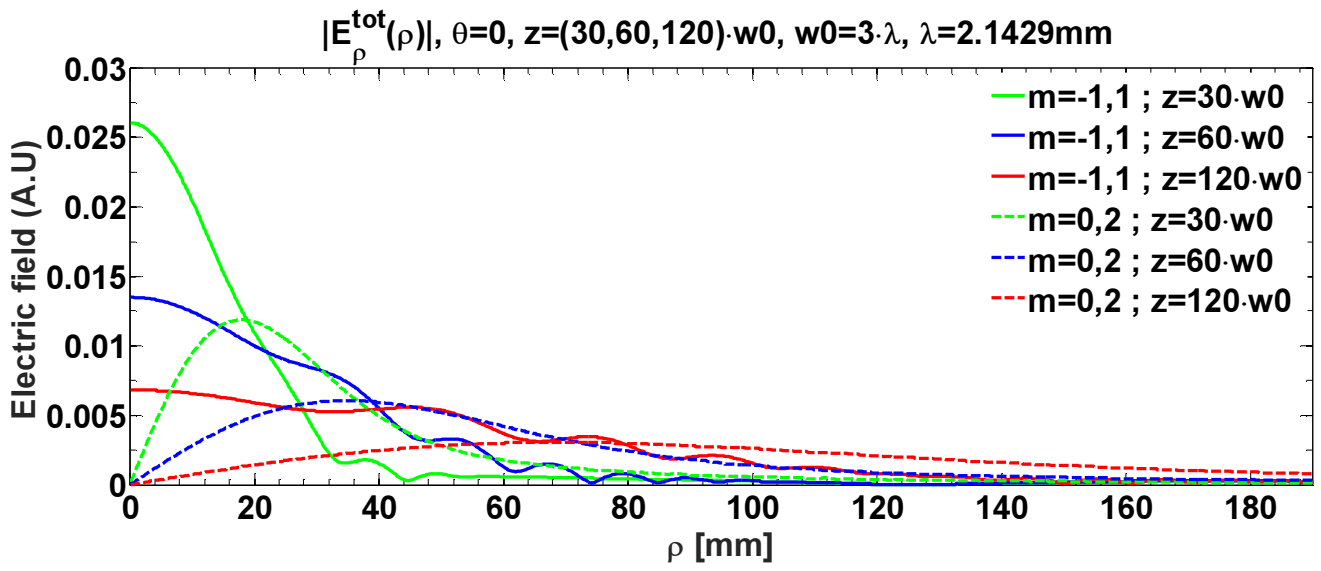

Fig.5. Linearly polarized Electric field distribution as a function of propagation distance (X-direction) at 140 GHz carrier frequency. The beam profile for different propagation distance (z-direction) is plotted.

The schematic of the proposed design for real-time channel de-multiplexing in spatial domain is shown in fig.6. Shifting the incoming THz beam spatially by 1 bit of the digital signal, we can effectively realize two independent data streams for the Bit error rate (BER) measurements. The phase plate (m=1) is placed in one of the channels and then both the independent channels are combined using the second beam splitter. The detector antenna is sensitive to one polarization alone.

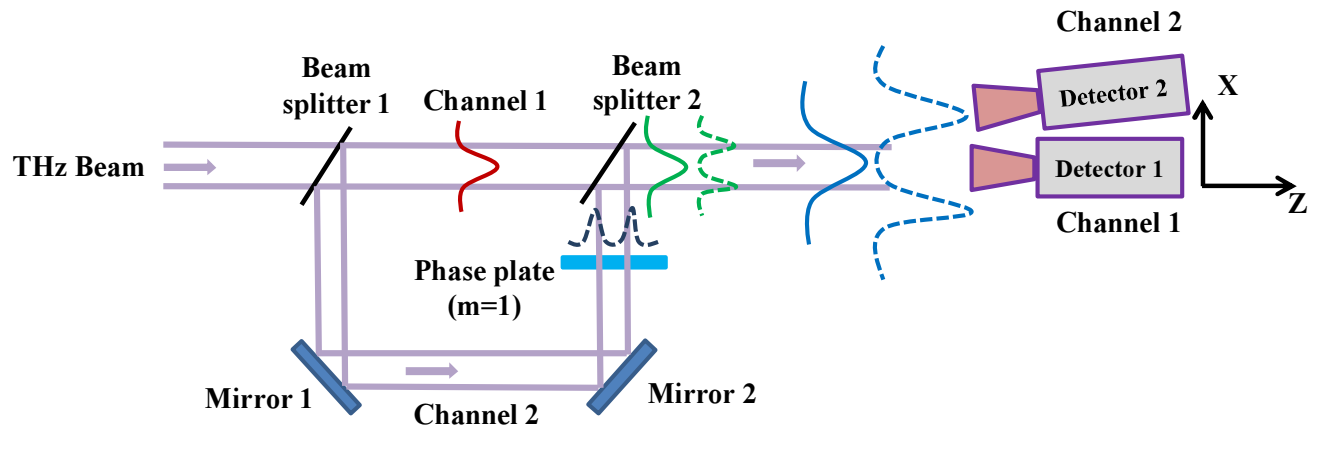

Fig.6. Schematic of the setup for multiplexing and spatial de-multiplexing of several wireless channels for THz communications

The electric field intensity of the channel 1 is maximum and channel 2 is minimum at X=0, so that the channel 1 can be de-multiplexed from the incoming multiplexed signal. Again, as moving the detector antenna along the X-direction, the electric field intensity of the channel 1 drops to a minimum value where we can de-multiplex channel 2. Thus, an OAM beams can be spatially de-multiplexed without using any phase plate of corresponding negative charge for de-multiplexing in the receiver end.

To conclude, we have presented the design for the generation of OAM using patterned phase plate and experimentally validated using the THz-Time domain system. The advantages of using patterned phase plate over the conventional spiral phase plate is that, the OAM charges can be tuned by simply stacking the additional phase plate. Due to the flat surface of the patterned phase plate, the system design can be compact for practical applications. We have also discussed the theoretical simulation on the de-multiplexing of different OAM states spatially along with the schematic of the proposed experiment.